\documentclass[12pt]{article}
\usepackage{lettrine}
\RequirePackage{fix-cm}
\usepackage{arxiv}
\usepackage[utf8]{inputenc} 
\usepackage[T1]{fontenc}    
\usepackage{hyperref}       
\usepackage{url}            
\usepackage{booktabs}       
\usepackage{amsfonts}       
\usepackage{nicefrac}       
\usepackage{microtype}      
\usepackage{lipsum}
\usepackage{graphicx}
\usepackage{pdflscape}
\usepackage{natbib}
\usepackage{fancyhdr}
\usepackage{multirow}
\usepackage{xcolor}
\graphicspath{ {./images/ } }
\fancypagestyle{mystyle}{
  \fancyhead{}
}
\title{Personality Trait Inference via Mobile Phone Sensors: A Machine Learning Approach}

\author{
    Wun Yung Shaney Sze \\
    Barcelona School of Economics \\
    Barcelona, Spain \\
    \texttt{wun.sze@bse.eu}
    \And
    Maryglen Pearl Herrero \\
    Barcelona School of Economics \\
    Barcelona, Spain \\
    \texttt{maryglen.herrero@bse.eu}
    \And
    Roger Garriga \\
    Koa Health \\
    Barcelona, Spain \\
    \texttt{roger.garrigacalleja@koahealth.com}
}

\begin{document}
\twocolumn[
\maketitle
\begin{abstract}
This study provides evidence that personality can be reliably predicted from activity data collected through mobile phone sensors. Employing a set of well-informed indicators calculable from accelerometer records and movement patterns, we were able to predict users' personality up to a 0.78 F1 score on a two-class problem. Given the fast growing number of data collected from mobile phones, our novel personality indicators open the door to exciting avenues for future research in social sciences.  Our results reveal distinct behavioral patterns that proved to be differentially-predictive of big five personality traits. They potentially enable cost-effective, questionnaire-free investigation of personality-related questions at an unprecedented scale. Overall, this paper shows how a combination of rich behavioral data obtained with smartphone sensing and the use of machine learning techniques can help to advance personality research and can inform both practitioners and researchers about the different behavioral patterns of personality. These findings have practical implications for organizations harnessing mobile sensor data for personality assessment,  guiding the refinement of more precise and efficient prediction models in the future.
\vspace{12pt}
\end{abstract}
] 

\subsubsection*{Introduction}
Is there a way for us to predict the personality of a person without them having to take surveys?  How much can one know about your personality type just by looking at the way you use your phone? Psychologists have widely adopted the Big Five Model, a framework encompassing the fundamental personality dimensions of Extraversion, Agreeableness, Conscientiousness, Neuroticism, and Openness. These dimensions capture traits such as gregariousness, trust, competence, emotional stability, and curiosity, respectively \citep{goldberg1992}. Traditionally, personality assessments relied on survey questionnaires for quantifying these traits based on individuals' responses. However, recent research demonstrates that digital behavior traces, including those generated through social media and smartphone usage, can effectively infer an individual's standing on these personality dimensions \citep{kosinski2013, azucar2018, park2018}.  The emerging field extends beyond personality trait prediction, encompassing the prediction of mental health states through mobile phone sensors, underscoring the utility of such data in enhancing our understanding and support of individuals' psychological well-being \citep{saeb2016, stachl2019}.

Traditional self-reported personality prediction has its limitations and can be time-consuming and inaccurate. Technological advances in mobile phones and sensing technology have now created the possibility to automatically record large amounts of data about humans’ natural behavior \citep{chittaranjan2013, farrahi2010, khwaja2019, montag2016, quercia2011}. The collection and analysis of these records makes it possible to analyze and quantify behavioral differences at unprecedented scale and efficiency.  The idea of predicting people's personalities from their mobile phone data stems from recent advances in data collection, machine learning, and computational social science showing that it is possible to infer various psychological states and traits from the way people use everyday digital technologies.

Exploration of smartphone data's potential in predicting the Big Five personality traits has been somewhat limited \citep{de2013, montag2016, schoedel2018}. Earlier studies reported relatively high predictive accuracy for these traits but with the constraint of small sample sizes \citep{chittaranjan2013, de2013}. However, subsequent research, conducted with more extensive participant pools, unveiled diminished predictive performance, exposing prior over-optimism attributable to model overfitting \citep{schoedel2018}. It's noteworthy that past investigations predominantly focused on communication-related behavior as predictors. Yet, smartphones encompass a multitude of functions, and personality traits can manifest in a broader array of behaviors \citep{funder2001, stachl2019}. Therefore, exploring behavioral patterns across various activities may extend predictive capabilities beyond just Extraversion, aligning with insights gleaned from social media data research \citep{schwartz2013, azucar2018}.

\begin{figure*}[!htbp]
    \centering
    \includegraphics[width=1.0\linewidth]{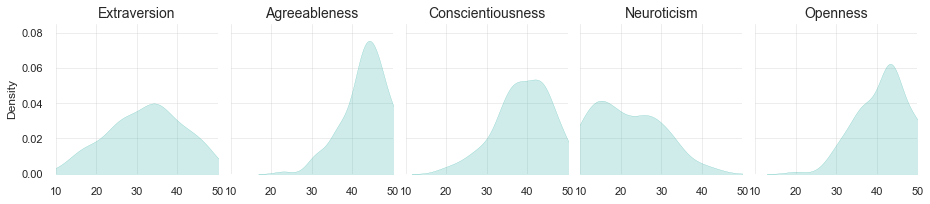}
    \caption{Distribution of the Big Five Personality Traits in the study population.}
\end{figure*}

Our objective twofold: (1) explore the relationship between personality and human behaviours sensed passively by smartphone sensors; (2) build predictive models based on previous literature and investigate how they perform over time and in different populations.  In our study, we employed a machine learning approach to predict the predictability of self-assessments for the Big Five personality traits using smartphone-derived variables, building upon previous research in this emerging field \citep{chittaranjan2013, de2013, schoedel2018}. Leveraging data collected from 144 participants through smartphone sensors, we derived a comprehensive set of behavioral features, encompassing aspects such as physical activity and daily behavioral patterns. Our research aims to expand the current understanding of the practical applications of smartphone data in personality trait prediction, while also considering the potential influence of cultural variations \citep{markus1991, terracciano2005}. 

\subsubsection*{Methods}
In this study, we first extract behavioral predictors from a diverse set of daily activities. Second, we use these variables to apply a machine learning approach for forecasting individuals' self-reported big five personality scores, encompassing both overarching factor levels and more specific facets.
Third, we analyze our results by looking at which features play a significant role in predicting each distinct dimension of an individual's personality traits. Finally, we discuss the impact of these variables within the context of prior research findings and outline potential avenues for further confirmatory investigations.

\paragraph{Dataset.}
The research dataset encompasses 3,282 recorded events spanning from March 2021 to May 2021, involving 144 distinct users who are all students at the London School of Economics (LSE). The data was exclusively gathered from Iphone devices. This dataset contains information on the activities in which users were involved and the duration of their engagement in these activities. The activities encompass five categories: walking, running, cycling, driving, and stationary periods. Additionally, the dataset includes details about the distance covered by the users each day during the recorded time frame, the number of floors they ascended and descended, and the longest period their phones remained untouched.

Furthermore, the dataset encompasses the participants' responses to the Big Five Inventory (BFI), a comprehensive questionnaire that assessed their Big Five personality traits \citep{john1991}. This self-report tool comprises 50 items, with responses rated on a 5-point Likert-type scale. The BFI is a well-established instrument widely employed in personality research, known for its strong psychometric properties \citep{mccrae1997, sutin2016}. Figure 1 provides a visual representation of the distributions of the five personality traits—Extraversion, Agreeableness, Conscientiousness, Neuroticism, and Openness—across our study cohort, using kernel density estimate plots.

\paragraph{Prediction Target.}
Our primary objective was to build predictive models for assessing the Big Five personality traits in users. We divided the classification into two parts: a 2-class classification and a 3-class classification. The former aimed to distinguish between two levels within each trait, while the latter categorized individuals into three levels.
We determine percentiles individually for each of the 5 personality target variable (representing personality traits). To create a 2-class, or binary classification, we set the threshold at the 50\% percentile for each variable. This categorizes values into two groups: 0 for users with variable values below the 50\% percentile and 1 for users with variable values above the 50\% percentile. In essence, this process transforms the target variable into a binary format by labeling values in the lower half (0-50\% percentile) as 0 and values in the upper half (50-100\% percentile) as 1.

In the context of a 3-class, or multiclass problem, we segment the values within each target variable into three distinct categories. We set the percentiles at 33\% and 67\% to create the 3-class classification. Values below the 33\% percentile are assigned the label 0, representing traits in the lower third, values between the 33\% and 67\% percentiles receive the label 1, indicating traits in the middle third, and values above the 67\% percentile are labeled as 2, signifying traits in the upper third. A similar approach has been utilized in various other personality prediction work using machine learning \citep{lima2014, teli2023}. 

\paragraph{Feature Extraction.}
Out of the initial dataset, we derived features extracted from physical activity data. This data encompasses information obtained from accelerometer data and processed in the phone. In particular, it records the specific activity a participant was engaged in at any given time, such as being stationary, walking, running, driving (automotive), or cycling, as well as the distance covered and floors ascended. This information allows us to infer various patterns, such as identifying extended periods of stationary time, which can be indicative of sleep duration if those happen at night. \\

The features were built following these steps:
\begin{itemize}
    \item First we built two synthetic activities. A physical activity label, that account for those instances where the user was either running or cycling, and a non-physical label for the rest of the instances. 
    \item Then, we aggregated the data in a daily basis. This aggregation included  the total amount of time a user was doing each activity (including the synthetic physical and non-physical activities), the number of occurances of each activity, the total distance covered, the floors ascended and the inferred hours of sleep.
    \item The features were subsequently categorized into two distinct groups: weekdays and weekends. This separation enables us to differentiate and capture behavioral variations that occur during typical working days and leisurely weekends, providing a more comprehensive understanding of participants' activity across different contexts and routines.
    \item Finally, we computed the average over weekdays, weekends and overall for each of the variables to summarize the entire history of the user. 
\end{itemize}

\begin{figure*}[!htbp]
    \centering
    \includegraphics[width=1.0\linewidth]{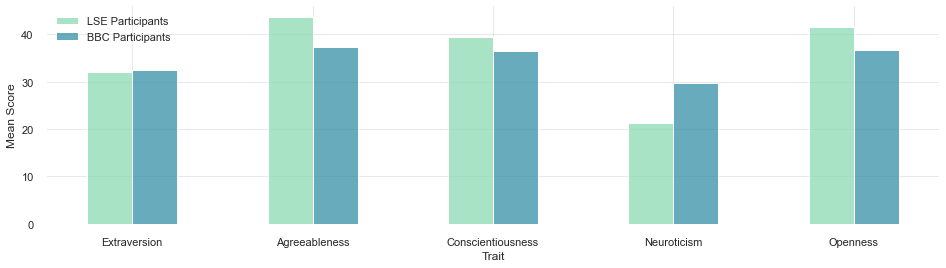}
    \caption{Comparison of Personality Traits against BBC Test.}
\end{figure*}

\paragraph{Feature Selection.}
After the feature extraction step, we obtained 83 features, a relatively high dimensional dataset given the size of our data. To reduce the dimension, we used recursive feature elimination with cross-validation (RFE-CV) to select the features \citep{darst2018}. The RFE-CV approach systematically eliminates less important features from a high-dimensional dataset by training a model, ranking feature importance, and removing the lowest-ranked features iteratively. For both the Random Forest and XGBoost models, and for each of the traits the feature selection process is carried out separately, resulting in distinct sets of features optimized for each of these models. For each model, RFE-CV automatically selects the number of features that resulted in the best mean score. 

\paragraph*{Machine Learning Models.}
To characterize the user and predict personality traits, we employed two main types of classification approaches depending on the prediction target: binary and multi-class. For each trait and modeling approach, we built different Machine Learning models. Given the relatively high interpretability and the high performance that tree-based models and boosting algorithms show in tabular data \citep{grinsztajn:2022:tree}, we chose two tree-based models, Random Forest and XGBoost \citep{chen:2016:xgboost}. \\

To assess the model's performance, we employed stratified k-fold cross-validation and computed the F1 scores, which combines both precision and recall. This involved dividing the dataset into k subsets or folds while ensuring that each fold maintained the same distribution of class labels as the original dataset. During each iteration, the model was trained on k-1 folds and evaluated on the remaining fold. This process was repeated k times, with each fold serving as the validation set exactly once. Using stratified k-fold is particularly addressing potential issues related to class imbalance. It guards against overfitting and ensures that the evaluation results are not skewed by the peculiarities of a single train-test split, especially when dealing with datasets where certain classes may be underrepresented.

We integrated Bayesian search techniques into our modeling pipeline for hyperparameter tuning \citep{perrone2019, lei2021, yang2007, stachl2019}. Bayesian optimization operates on the principle of sequentially exploring and exploiting the hyperparameter space to identify the configuration that optimizes a chosen evaluation metric, such as F1 score in our case. This process begins with an initial set of hyperparameter configurations, often determined through random or grid search. Subsequently, Bayesian optimization refines this initial set by iteratively selecting new configurations based on their predicted performance using a surrogate probabilistic model.

The Bayesian search systematically explored the hyperparameter space by selecting candidate configurations that showed promise in improving model performance \citep{snoek2012}. It utilized a probabilistic surrogate model to model the relationship between hyperparameters and the chosen evaluation metric, effectively guiding the search towards the most promising regions of the hyperparameter space.

\subsubsection*{Results}
\paragraph{Population Comparison.}
Our study cohort differs from the general population. Figure 1 shows that the distributions of all personality traits except Extraversion are skewed. We compared the statistics of average traits of our study cohort to the Big Personality Test dataset conducted by the British Broadcasting Corporation (BBC). The BBC test uses a sample of $N = 386,375$ British residents, mapping the distributions of the Big Five Personality traits \citep{rentfrow2015}. 

When juxtaposed with the general British population, shown in Figure 2, LSE students exhibited quantifiable variations. Specifically, LSE students demonstrated significantly higher levels of Openness, with a mean score of 41.51 compared to the BBC dataset's mean of 36.7. This suggests a propensity for greater receptivity to diverse academic and cultural influences within the LSE academic environment. Additionally, our analysis revealed that LSE students tend to score lower in Neuroticism, recording an average of 21.27 compared to the BBC dataset's mean of 29.7. This finding implies a potential inclination toward resilience and adaptability fostered by their academic pursuits. Moreover, our cohort displayed higher levels of Agreeableness, registering an average score of 43.74, in contrast to the BBC dataset's mean of 37.4. This observation implies a greater propensity for cooperative and harmonious social interactions, possibly influenced by their academic and communal experiences. These empirically observed distinctions underscore the unique character of our study cohort and provide crucial context for our ongoing research into predicting Big Five personalities from mobile phone sensor data within the specific academic setting of the cohort.

\paragraph{Model performance.}
We present the results of the binary and multiclass classification in Table 1. Considerable differences in the F1 scores across various personality traits are observed. Specifically, the binary models exhibit notable performance ranging between 0.56 and 0.78 in the all traits. The highest performance is achieved when predicting Extraversion, with an F1 score of 0.78. Interestingly, Random Forest showed higher performance in Extraversion, Agreebleness and Neuroticism, while XGBoost performed better for Conscientiousness and Openness.

The multiclass models exhibit a more varied performance ranging between 0.25 and 0.47 in the all traits. The highest performance is achieved when predicting Openness, with an F1 score of 0.47. For multiclass models, Random Forest showed higher performance in predicting Neuroticism, XGBoost performed better for Agreebleness, Conscientiousness, Openness, and both models scored about the same for predicting Extraversion. 

\begin{table*}[!htbp]
\centering
\renewcommand{\arraystretch}{1.25} 
\begin{tabular}{|c|c|c|c|c|}
\hline
 & \multicolumn{2}{c|}{\textbf{Binary}} & \multicolumn{2}{c|}{\textbf{Multiclass}} \\
\cline{2-5} 
\textbf{Personality trait} & \textbf{Random Forest} & \textbf{XGBoost} & \textbf{Random Forest} & \textbf{XGBoost} \\
\hline
\textbf{EXT} & 0.78 & 0.61 & 0.39 & 0.39 \\
\hline
\textbf{AGR} & 0.58 & 0.56 & 0.36 & 0.42 \\
\hline
\textbf{CON} & 0.64 & 0.75 & 0.25 & 0.33 \\
\hline
\textbf{NEU} & 0.61 & 0.58 & 0.47 & 0.36 \\
\hline
\textbf{OPE} & 0.58 & 0.61 & 0.42 & 0.47 \\
\hline
\end{tabular}
\caption{Cross validated F1 Scores for out-of-sample classifications.}
\end{table*}

\paragraph{Feature Analysis.}
To obtain the top predictive features for each trait, individual models were trained and tested using the dataset split by binary or multiclass levels. Table A1 in the appendix presents the top three features for each model associated with a particular personality trait. 

\textit{Extraversion} was predicted by the time spent stationary and doing outdoor activities, such as automotive (car), running or cycling. Extraverts are known for their sociable and outgoing nature, and this result aligns with the idea that they may engage in more activities outside their home or workplace, leading to reduced stationary time.

\textit{Agreeabless} was predicted by the ascended and descended floors, average active pace and number of accumulated steps, this may imply that agreeable individuals may engage more in physical activities like walking and stair climbing due to their health-conscious nature, preference for routine, and tendency towards social interactions.

\textit{Conscientiousness} was predicted by the sleep related features, such as time of waking up, cycling or accumulated steps during the weekend. This may imply that conscientious individuals are more likely to wake up at certain hours and maintain their physical activity routines even during weekends, reflecting their self-discipline, organization, and commitment to personal goals and health.

\textit{Neuroticism} was predicted by the number of floors ascended and descended during the weekday, distance travelled and the duration that one takes part in physical activities. This may suggest that individuals with higher levels of Neuroticism might use physical activities, like ascending and descending stairs during weekdays, as a way to manage stress and anxiety, or it could reflect their varied response to daily routines and stressors.

\textit{Openness} was predicted by amount of sleep that a user gets and cycling during the weekends. This may indicate that individuals high in Openness, who are often curious and open to new experiences, might prioritize sufficient sleep for cognitive and creative functioning and engage in exploratory activities, leading to traveling in bicycle during the weekend.

\paragraph{Discussion.}
Our study culminates in several noteworthy findings. Primarily, our results demonstrate the potential of leveraging activity data collected through mobile phone sensors for classifying the Big Five personality traits of students with a with a performance between 0.56 and 0.78 for F1 scores. This underscores the viability of utilizing pervasive technology as a conduit to gain insights into individuals' psychological dispositions \citep{wu2015}. In an era where smartphones have become ubiquitous companions, the ability to discern personality traits through unobtrusive data collection is a paradigm-shifting advancement in the field of personality psychology.

Each personality trait's prediction was tied to distinct behavioral patterns observed through the sensors. The discovery that Extraversion is best predicted by the time spent in outdoor activities substantiates the theory that extraverts tend to be more sociable and outgoing, possibly engaging in more external activities, thereby reducing the time spent in stationary states \citep{srivastava2008, lochbaum2013}. Meanwhile, the linkage between Agreeableness and the number of steps and floors taken highlights the physical aspect of Agreeableness. This suggests that individuals who exhibit more physical activity, such as stair climbing, tend to possess higher levels of Agreeableness, a facet of their personality possibly reflected in their willingness to engage in physical cooperation.

The relationship between Conscientiousness and the number of steps and and cycling duration on the weekends can imply that individuals high in Conscientiousness may engage more actively in physical activities during their leisure time, reflecting their self-discipline and commitment to personal health and goals, even outside of structured weekday routines. Neuroticism was best predicted by floors descended or ascended on the weekdays, and Openness was best predicted by the sleep duration and cycling during the weekends.

Our research, confined to analyzing activity data from smartphone sensors within a student population, opens up novel avenues for understanding personality traits. These insights are pivotal in shaping interventions and recommendations for personal development and well-being, tailored to individual lifestyle choices and personality profiles.

\paragraph{Further Research.}
Despite the valuable insights offered by our study, it is imperative to acknowledge inherent limitations. Our dataset primarily comprises students from LSE, thereby prompting questions about the generalizability of our findings to a more diverse and representative population. As a plausible remedy, forthcoming research endeavors should prioritize expanding the sample size to encompass a broader and more varied demographic spectrum, thereby ensuring that the derived insights retain their applicability across heterogeneous populations.

In light of the observed comparability discrepancies between our university student cohort and the general population, as discerned through comparative analyses with data compiled by the BBC, an in-depth exploration into the demographic biases embedded within our dataset becomes imperative. Further investigations should strive to unravel how the distinctive characteristics of our cohort potentially influence the predictive accuracy of personality traits. Consequently, there exists an indispensable need to engage in a comprehensive exploration of this domain, elucidating the boundary conditions and extending the domain of generalizability pertaining to our findings.

It is noteworthy to consider the rich tapestry of analogous methodologies and techniques that have been deployed within the landscape of personality prediction. These encompass the utilization of artificial neural network models for classification \citep{basaran2021}, the employment of textual data and their alignment with prevailing personality models \citep{kunte2019, christian2021}, as well as the innovative approach of harnessing graphology and digital handwriting analysis to discern personality traits \citep{samsuryadi2023}. The integration of these diverse methodologies promises an enhanced understanding of human personality traits.

Our study not only underscores the vast potential inherent in mobile phone sensor data for personality trait prediction but also casts a luminous spotlight on the intricate nature of the relationship between personality and lifestyle. This, in turn, lays the cornerstone for an auspicious avenue of research, geared towards further enriching our understanding of the intricate nexus in the field of personality prediction. 

\subsubsection*{Acknowledgements}
The authors would like to thank Koa Health for providing the dataset and laying the groundwork for this line of inquiry.  The authors would also like to thank Hannes Mueller, Jesús Cerquides, Christian Brownless, and Elliot Motte for their encouragement in this intellectual
pursuit.

\bibliography{main}
\bibliographystyle{apalike}
\nocite{*}

\newpage
\clearpage
\appendix
\renewcommand{\thefigure}{A\arabic{figure}}
\renewcommand{\thetable}{A\arabic{table}}
\setcounter{figure}{0}  
\setcounter{table}{0}

\section{Appendix}
\begin{minipage}{\textwidth}
    \centering
    \includegraphics[width=1.0\linewidth]{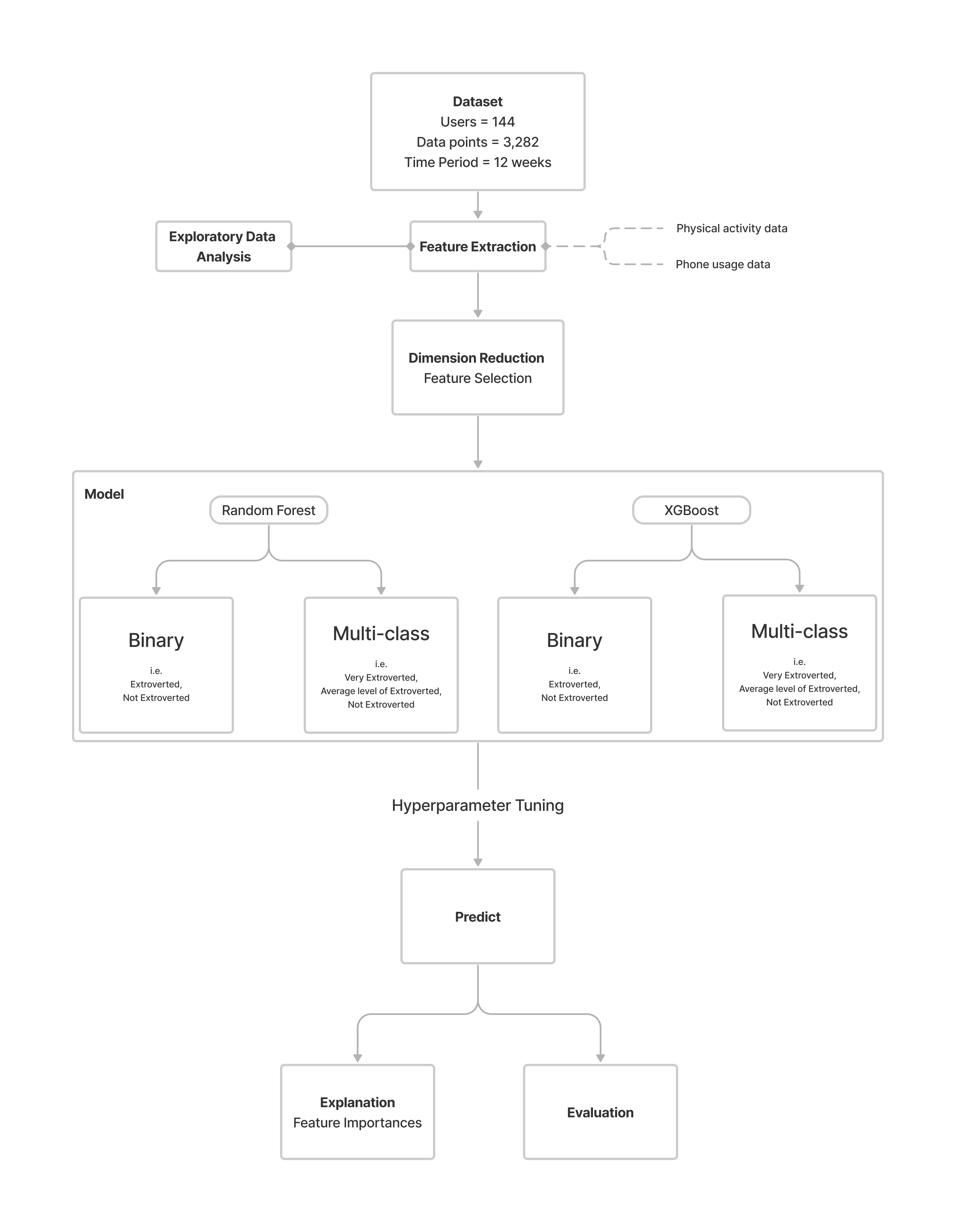}
\end{minipage}

\begin{table*}[!htbp]
\centering
\footnotesize
\setlength{\tabcolsep}{4pt}
\renewcommand{\arraystretch}{1.7}
\begin{tabular}{ccc|ccc|c}
\cline{4-6}
&                                                              &     & \multicolumn{3}{c|}{\textbf{Most Important Features}}                                                                                                               &  \\ \cline{4-6}
&                                                              &     & \multicolumn{1}{c|}{\textbf{1st}}                                & \multicolumn{1}{c|}{\textbf{2nd}}                      & \textbf{3rd}                            &  \\ \cline{1-6}
\multicolumn{1}{|c|}{\multirow{10}{*}{\textbf{Binary}}}     & \multicolumn{1}{c|}{\multirow{5}{*}{\textbf{RF}}} & EXT & \multicolumn{1}{c|}{Stationary Duration weekday}                 & \multicolumn{1}{c|}{Automotive Count weekday}          & Automotive Duration weekday             &  \\
\multicolumn{1}{|c|}{}                                      & \multicolumn{1}{c|}{}                                        & AGR & \multicolumn{1}{c|}{Floors Ascended weekend}            & \multicolumn{1}{c|}{Floors Descended weekend} & Physical Activity Count weekday          &  \\
\multicolumn{1}{|c|}{}                                      & \multicolumn{1}{c|}{}                                        & CON & \multicolumn{1}{c|}{Hour of Waking Up weekday}             & \multicolumn{1}{c|}{Automotive Count weekday}          & Physical Activity Count weekend &  \\
\multicolumn{1}{|c|}{}                                      & \multicolumn{1}{c|}{}                                        & NEU & \multicolumn{1}{c|}{Floors Ascended weekend}            & \multicolumn{1}{c|}{Stationary Duration weekday}       & Distance Travelled weekday              &  \\
\multicolumn{1}{|c|}{}                                      & \multicolumn{1}{c|}{}                                        & OPE & \multicolumn{1}{c|}{Sleep Duration weekend}             & \multicolumn{1}{c|}{Sleep Duration weekday}            &      Stationary Duration weekday        &  \\ \cline{2-6}
\multicolumn{1}{|c|}{}                                      & \multicolumn{1}{c|}{\multirow{5}{*}{\textbf{XGB}}}       & EXT & \multicolumn{1}{c|}{Accumulated Steps weekend}          & \multicolumn{1}{c|}{Floors Ascended weekday}           & Automotive Duration weekday           &  \\
\multicolumn{1}{|c|}{}                                      & \multicolumn{1}{c|}{}                                        & AGR & \multicolumn{1}{c|}{Floors Ascended weekend}            & \multicolumn{1}{c|}{Stationary Count weekday}          & Sleep Duration weekday                  &  \\
\multicolumn{1}{|c|}{}                                      & \multicolumn{1}{c|}{}                                        & CON & \multicolumn{1}{c|}{Cycling Duration weekday}                    & \multicolumn{1}{c|}{Stationary Count weekend} & Running Duration weekday                &  \\
\multicolumn{1}{|c|}{}                                      & \multicolumn{1}{c|}{}                                        & NEU & \multicolumn{1}{c|}{Physical Activity Duration weekend} & \multicolumn{1}{c|}{Stationary Duration weekday}       & Cycling Duration weekday                &  \\
\multicolumn{1}{|c|}{}                                      & \multicolumn{1}{c|}{}                                        & OPE & \multicolumn{1}{c|}{Automotive Count weekday}                    & \multicolumn{1}{c|}{Cycling Duration weekday}          & Running Duration weekend       &  \\ \cline{1-6}
\multicolumn{1}{|c|}{\multirow{10}{*}{\textbf{Multiclass}}} & \multicolumn{1}{c|}{\multirow{5}{*}{\textbf{RF}}} & EXT & \multicolumn{1}{c|}{Stationary Duration Weekday}                 & \multicolumn{1}{c|}{Automotive Count weekday}          & Distance Travelled weekday             &  \\
\multicolumn{1}{|c|}{}                                      & \multicolumn{1}{c|}{}                                        & AGR & \multicolumn{1}{c|}{Activity Count for 24h weekday}              & \multicolumn{1}{c|}{Sleep Duration weekend}   & Floors Ascended weekend         &  \\
\multicolumn{1}{|c|}{}                                      & \multicolumn{1}{c|}{}                                        & CON & \multicolumn{1}{c|}{Hour of Waking Up weekday}                   & \multicolumn{1}{c|}{Hour of Asleep weekend} & Sleep Duration weekend          &  \\
\multicolumn{1}{|c|}{}                                      & \multicolumn{1}{c|}{}                                        & NEU & \multicolumn{1}{c|}{Distance Travelled weekday}                  & \multicolumn{1}{c|}{Floors Descended weekday}          & Stationary Duration weekday              &  \\
\multicolumn{1}{|c|}{}                                      & \multicolumn{1}{c|}{}                                        & OPE & \multicolumn{1}{c|}{Stationary Duration weekday}                 & \multicolumn{1}{c|}{Sleep Duration weekday}            & Sleep Duration weekday                  &  \\ \cline{2-6}
\multicolumn{1}{|c|}{}                                      & \multicolumn{1}{c|}{\multirow{5}{*}{\textbf{XGB}}}       & EXT & \multicolumn{1}{c|}{Running Duration weekend}           & \multicolumn{1}{c|}{Stationary Duration weekday}       & Cycling Duration weekday                 &  \\
\multicolumn{1}{|c|}{}                                      & \multicolumn{1}{c|}{}                                        & AGR & \multicolumn{1}{c|}{Floors Descended weekend}           & \multicolumn{1}{c|}{Running Count weekday}             & Floors Ascended weekday                &  \\
\multicolumn{1}{|c|}{}                                      & \multicolumn{1}{c|}{}                                        & CON & \multicolumn{1}{c|}{Cycling Count weekend}              & \multicolumn{1}{c|}{Sleep Duration weekend}   & Accumulated Steps weekend      &  \\
\multicolumn{1}{|c|}{}                                      & \multicolumn{1}{c|}{}                                        & NEU & \multicolumn{1}{c|}{Cycling Duration weekend}           & \multicolumn{1}{c|}{Walking Count weekend}    & Walking Duration weekday               &  \\
\multicolumn{1}{|c|}{}                                      & \multicolumn{1}{c|}{}                                        & OPE & \multicolumn{1}{c|}{Cycling Duration Pct weekend}       & \multicolumn{1}{c|}{Cycling Duration weekend} & Cycling Count weekend          &  \\ \cline{1-6}
\end{tabular}
\caption{Most important features for binary and multiclass classifications.}
\end{table*}
\end{document}